\documentclass[conference]{IEEEtran}
\usepackage{cite}

\usepackage[printonlyused]{acronym} 

\usepackage[cmex10]{amsmath} 
\usepackage{amsthm} 
\usepackage{amssymb} 
\usepackage{array}
\usepackage{cancel} 
\usepackage[keeplastbox]{flushend}
\usepackage[caption=false,font=footnotesize]{subfig}
\usepackage[pdftex]{graphicx}
\graphicspath{{./figs/}}
\DeclareGraphicsExtensions{.pdf,.jpeg,.png}

\usepackage{color}

\usepackage{balance} 
\usepackage{bm} 

\usepackage{threeparttable} 
\usepackage{multicol} 
\usepackage{multirow} 
\usepackage{rotating} 

\usepackage{upgreek} 
\usepackage{url} 
\usepackage{verbatim} 

\usepackage{mathtools} 
\usepackage{mathrsfs} 

\usepackage{xfrac} 

\usepackage{optidef}
\usepackage{fancyhdr}

\newcommand{\HypTst}[2]{\raise2.5ex\hbox{\scriptsize$#1$} \hspace{-1.3em}\displaystyle\gtreqless\hspace{-1.2em}\raise-1.1em\hbox{\scriptsize$#2$}}

\newcommand{\cref}[1]{C\ref{#1}}

\acrodef{3-G}{3-Generation}
\acrodef{3GPP}{3rd Generation Partnership Project}
\acrodef{a.k.a}{also-known-as}
\acrodef{AAS}{active antenna system}
\acrodef{ADMM}{alternating direction method of multipliers}
\acrodef{AoA}{angle-of-arrival}
\acrodef{AoD}{angle-of-departure}
\acrodef{AHB}{Abel hybrid bound}
\acrodef{A-TS}{alternating Taylor\'s series}
\acrodef{A-LASSO}{\emph{adaptive}-least absolute shrinkage and selection operator}
\acrodef{AWGN}{Additive White Gaussian Noise} 
\acrodef{AGC}{Automated Gain Control}
\acrodef{BER}{Bit Error Rate}
\acrodef{BB}{Bhattacharyya Bound}
\acrodef{BF}{beamformer}
\acrodef{BFGS}{Broyden-Fletcher-Goldfarb-Shanno}
\acrodef{BS}{base-station}
\acrodef{cdf}{cumulative distribution function}
\acrodef{CEP}{Circular Error Probability}
\acrodef{CoMP}{Coordinated Multi-Point}
\acrodef{CIR}{Channel Impulse Response}
\acrodef{C-RDM}{Cross-Ranging Direction Matrix}
\acrodef{C-NLS}{Constrained Non-linear Least Squares}
\acrodef{CRLB}{Cram\'{e}r-Rao Lower Bound}
\acrodef{CSIT}{Channel State Information at the Transmitter}
\acrodef{CP}{Cyclic-Prefix}
\acrodef{DC}{Distance Contraction}
\acrodef{DCT}{Distance Contraction Theory}
\acrodef{DE}{Distance Error}
\acrodef{DFT}{Discrete Fourier Transform}
\acrodef{DL}{Downlink}
\acrodef{DoA}{Direction-of-Arrival}
\acrodef{DSSS}{Direct Sequence Spread Spectrum}
\acrodef{EDF}{Euclidean Distance Function}
\acrodef{EDM}{Euclidean Distance Matrix}
\acrodef{EFIM}{Equivalent Fisher Information Matrix}
\acrodef{EHF}{Extremely High Frequency}
\acrodef{EKT}{Euclidean Kernel Transformation}
\acrodef{EK}{Euclidean Kernel}
\acrodef{ERII}{Equivalent Ranging Information Intensity}
\acrodef{ETSI}{European Telecommunications Standards Institute}
\acrodef{FIM}{Fisher Information Matrix}
\acrodef{FCC}{Federal Communications Commission}
\acrodef{FDD}{Frequency-Division-Duplex}
\acrodef{GLE}{Geometric-constrained Location Estimation}
\acrodef{G-LS}{Global Least Squares}
\acrodef{GNSS}{Global Navigation Satellite System}
\acrodef{GP}{Gaussian Process}
\acrodef{GP-LVM}{Gaussian Process Latent Variable Model}
\acrodef{GPS}{Global Positioning System}
\acrodef{GDC}{Global Distance Continuation}
\acrodef{GDOP}{Geometric Dilution of Precision}
\acrodef{GTRS}{Generalized Trust Region Subproblems}
\acrodef{HCRB}{Hammersley-Chapmann-Robbins Bound}
\acrodef{HB}{Hybrid Bound}
\acrodef{HOSVD}{High-Order Singular-Value-Decomposition}
\acrodef{HPBW}{half power beamwidth}
\acrodef{iid}{independent identically distributed}
\acrodef{ICT}{Information Communication Technology}
\acrodef{I-EKT}{Inverse Euclidean Kernel Transformation}
\acrodef{IoT}{Internet-of-Things}
\acrodef{iff}{if and only if}
\acrodef{IFT}{Inverse Fourier Transform}
\acrodef{ITU}{International Telecommunication Union}
\acrodef{I/N}{Interference-to-Noise}
\acrodef{KKT}{Karush-Kuhn-Tucker}
\acrodef{KLT}{Karhunen-Lo\'{e}ve Transform}
\acrodef{LASSO}{least absolute shrinkage and selection operator}
\acrodef{LARSO}{Least Absolute Residual and Selection Operator}
\acrodef{LBSs}{Location Based Services}
\acrodef{LM}{Levenberg-Marquardt}
\acrodef{LQ}{Link-Quality}
\acrodef{LS}{Least Squares}
\acrodef{LLS}{Linearized Least Squares}
\acrodef{LT}{Location-Tracking}
\acrodef{LTE}{Long Term Evolution}
\acrodef{LOS}{line-of-sight}
\acrodef{LOESS}{Locally weighted Scatterplot Smoothing}
\acrodef{LP}{Linear Programming}
\acrodef{MAP}{Maximum A Posteriori}
\acrodef{MOO}{Multiple-Objective Optimization}
\acrodef{MOS}{Mean Opinion Score}
\acrodef{MIMO}{multiple-input-multiple-output}
\acrodef{MMSE}{Minimum Mean Squared Error}
\acrodef{mmW}{millimeter wave}
\acrodef{MSE}{Mean Square Error}
\acrodef{ML}{Maximum Likelihood}
\acrodef{MS}{mobile-station}
\acrodef{MDS}{Multidimensional Scaling}
\acrodef{MSPE}{Mean-Squared-Position-Error}
\acrodef{MUSIC}{MUltiple SIgnal Classification}
\acrodef{N-DC}{Negative-Distance Contraction}
\acrodef{NLOS}{non-line-of-sight}
\acrodef{NLS}{Non-Linear-Least Squares}
\acrodef{NP-hard}{Non-deterministic Polynomial-time hard}
\acrodef{NPRM}{Notice of Proposed Rulemaking}
\acrodef{OFDM}{orthogonal frequency-division multiplexing}
\acrodef{OMP}{Orthogonal Matching Pursuit}
\acrodef{OTDoA}{observed-Time-Difference-of-Arrival}
\acrodef{psd}{positive semi-definite}
\acrodef{PSD}{power spectral density}
\acrodef{pdf}{probability density function}
\acrodef{PDP}{Power Delay Profile}
\acrodef{P-DC}{Positive-Distance Contraction}
\acrodef{PE}{Position Error}
\acrodef{PEB}{Position Error Bound}
\acrodef{PREB}{position-rotation error bound}
\acrodef{PSM}{Positive Semi-definite Matrix}
\acrodef{POCS}{Projection On Convex Sets}
\acrodef{PCA}{Principal Component Analysis}
\acrodef{PPCA}{Probabilistic Principal Component Analysis}
\acrodef{QoD}{Quality-of-Design}
\acrodef{QoL}{Quality-of-Location}
\acrodef{QoS}{Quality-of-Service}
\acrodef{QoE}{Quality-of-Experience}
\acrodef{QoI}{Quality-of-Information}
\acrodef{R95}{95$\%$ Radius}
\acrodef{RBF}{Radial Basis Function}
\acrodef{RDM}{Ranging Direction Matrix}
\acrodef{REB}{Rotation Error Bound}
\acrodef{RF}{radio frequency}
\acrodef{R-GDC}{Range-Global Distance Continuation}
\acrodef{R-LS}{Regularized Least-Squares}
\acrodef{RSS}{Received Signal Strength}
\acrodef{RSSI}{Received Signal Strength Index}
\acrodef{RMB}{Reuven-Messer Bound}
\acrodef{RMSE}{root-mean-squared-error}
\acrodef{RII}{Ranging Information Intensity}
\acrodef{RR}{Ridge-Regression}

\acrodef{SCE}{Sparse Channel Estimator}
\acrodef{SCM}{spatial channel model}
\acrodef{SDP}{Semi Definite Programming}
\acrodef{SIMO}{Single-Input-Multiple-Output}
\acrodef{SISO}{Single-Input-Single-Output}
\acrodef{SLAM}{Simultaneous Localization and Mapping}
\acrodef{SMACOF}{Stress-of-a-MAjorizing-Complex-Objective-Function}
\acrodef{SQP}{Sequential Quadratic Programming}
\acrodef{SR-LS}{Squared-Range Least-Squares}
\acrodef{SR-GDC}{Square Range-Global Distance Continuation}
\acrodef{SER}{Symbol-Error-Rate}
\acrodef{SB}{Stochastic Bound}
\acrodef{SNR}{Signal-to-Noise Ratio}
\acrodef{SHF}{Super High Frequency}
\acrodef{SINR}{Signal-to-Interference-Noise Ratio}
\acrodef{SD}{Steepest-Descent}
\acrodef{SA}{Simulated-Annealing}
\acrodef{SR}{Squared Range}
\acrodef{TDoA}{Time-Difference-of-Arrival}
\acrodef{TS}{Taylor's Series}
\acrodef{ToF}{Time-of-Flight}
\acrodef{ToA}{Time-of-Arrival}
\acrodef{TDD}{Time-Division-Duplex}
\acrodef{TTI}{Transmission Time Interval}
\acrodef{ULA}{uniform linear array}
\acrodef{UMTS}{Universal Mobile Telecommunications System}
\acrodef{URA}{uniform rectangular array}
\acrodef{UWB}{Ultra-WideBand}
\acrodef{UHF}{Ultra High Frequency}

\acrodef{WLS}{Weighted Least Squares}
\acrodef{WC}{Weighted Centroid}

\acrodef{ZC}{Zadoff-Chu}
\acrodef{ZF}{Zero-Forcing}

\usepackage{xcolor}
\makeatletter
\let\old@ps@headings\ps@headings
\let\old@ps@IEEEtitlepagestyle\ps@IEEEtitlepagestyle
\def\confheader#1{%
\def\ps@headings{%
\old@ps@headings%
\def\@oddhead{\strut\hfill#1\hfill\strut}%
\def\@evenhead{\strut\hfill#1\hfill\strut}%
}%
\def\ps@IEEEtitlepagestyle{%
\old@ps@IEEEtitlepagestyle%
\def\@oddhead{\strut\hfill#1\hfill\strut}%
\def\@evenhead{\strut\hfill#1\hfill\strut}%
}%
\ps@headings%
}
\makeatother

\makeatletter

\begin{document}
%
\title{High Reliability Downlink MU-MIMO: New OSTBC Approach and Superposition Modulated Side Information}
\author{
\IEEEauthorblockN{Nora Boulaioune, Nandana Rajatheva, Matti Latva-aho}
\IEEEauthorblockA{Centre for Wireless Communications, 
University of Oulu, Finland\\
Email:\{nora.boulaioune, nandana.rajatheva, matti.latva-aho\}@oulu.fi}
}

\maketitle

\begin{abstract}
In this paper a proposal to improve the reliability of a downlink multiuser (MU) MIMO transmission scheme is investigated with the use of a new approach in orthogonal space-time block codes (OSTBC) and network coding with a superposition modulated system and side information. With the new encoded OSTBC approach, diversity is offered where each user receives all other users' symbols, which allows the recovery of symbols in several ways. In addition, multiple users can be accommodated with the same resource, which is quite useful in a wireless system where resources are always restricted. By employing superposition modulation, the side information needed for error recovery can be transmitted over the same resource used for the normal information frame. In addition, the proposed system exploits diversity through a novel technique of sub-constellation alignment-based signal combining for efficient side information dissemination. A detailed analysis of the new OSTBC approach is carried out. It is shown that the performance of the MU-MIMO system can be improved significantly in terms of block and frame error rates (BLER, FER) considered as reliability measures. By accommodating a reasonable number of multiple users, high reliability is achieved at the expense of the rate. To compensate for the low rate, conventional OSTBC can be considered and simulation results are shown, where, as a penalty to pay, multiple orthogonal resources are required.\\

\end{abstract}

\section{Introduction}

In a significant departure from LTE based 4G existing systems, the future communication networks will proliferate into many areas each with completely different set of requirements. While broadband is the mainstay initially resulting in enhanced mobile broadband (eMMB) category, pervasive nature of internet of things, sensors and machine type devices bring new challenges in the required quality of service (QoS) metrics. To handle this flood in data traffic and to improve system performance in various aspects, significant research endeavors have been concentrated towards advancement of 5G communication systems \cite{Mueck-5GCHAMPION-2016}.

With the advent of 5G, there has been considerable effort in communications with higher reliability. If one is to consider control applications in an industry setting and to be able to replace wires, a credible demonstration of achieved reliability through wireless is extremely important. One can easily think of factories of future (FoF), autonomous vehicles (AVs), surgeries with robotic equipment controlled from a distant place and so on. All these depend critically on the reliability of the link and its performance. In general one would immediately think of applying different methods of diversity techniques as discussed in \cite{johansson2015radio} where the best method to guarantee a higher state of transmission reliability is through diversity. Various methodologies for accomplishing high reliability in a wireless system are compared there.

In this paper \footnote{Based on the first authors Master thesis available at jultika.oulu.fi} , we investigate \footnote{Version available in arxiv 1911.05347} the use of side information (SI) assisted error recovery strategy like network coding. This is first examined in \cite{nazer2011reliable} where it is proved that it is possible to tackle interference and obtain more reliable communication over a network. This is with novel techniques adapted for mitigating wireless interference to achieve that objective through network coding. As mentioned in \cite{haghighat2017high}, a significant downside in the usage of network coding is the utilization of additional physical resources to provide SI dissemination. To circumvent this issue we also investigate the application of superposition modulation \cite{hoeher2011superposition} to guarantee the SI dissemination and to keep the network away from the overall spectral efficiency deterioration. The early work by Alamouti \cite{alamouti1998simple} considered the application of OSTBC to improve the diversity of wireless communication system. Later on, authors in \cite{tarokh1998application} investigated extended forms of OSTBC where the transmission scheme found by Alamouti is generalized to an arbitrary number of transmit antennas. Due to the special structure of orthogonal designs, these codes are able to accomplish full diversity and have a straightforward linear maximum-likelihood decoding algorithm at the receiver. In \cite{tarokh1999space}, the performance of the OSTBCs in \cite{tarokh1998application} is evaluated. The models of OSTBCs are presented with suitable encoding and decoding algorithms. 
The authors in \cite{su2004systematic} considered a systematic design to create high-rate OSTBCs from complex orthogonal designs for any number of transmit antennas where those have the best known rates. A similar orthogonal design for eight transmit antennas is presented in \cite{bao2009high}. 
\\

Motivated by the aforementioned issues, the purpose behind this paper is to improve and evaluate the reliability performance of a MU-MIMO downlink transmission scheme. In contrast to previous work, our research objective is to combine side information (SI) with superposition modulation and the proposed new encoded OSTBC approach, where multiple users share the same resource. In addition to the diversity offered by OSTBC, the proposed design also exploits diversity through a novel strategy of constellation alignment based signal combining to enhance reliability of the SI transmission.

The reminder of the paper is organized as follows. First, in Section \ref{sec:model}, the general system model is described as illustrated in Fig.\ref{fig:System-Model}. In Section \ref{sec:High-Reliable XOR and OSTBC Approach}, details of the transmitter and receiver processing are given. In Section \ref{sec:results}, simulation results are provided and discussed. Finally, in Section \ref{sec:Conclusion}, our conclusion and comments are presented.

\section{System Model}
\label{sec:model}
\begin{figure}[h]
\centering
\includegraphics[width=0.9\columnwidth]{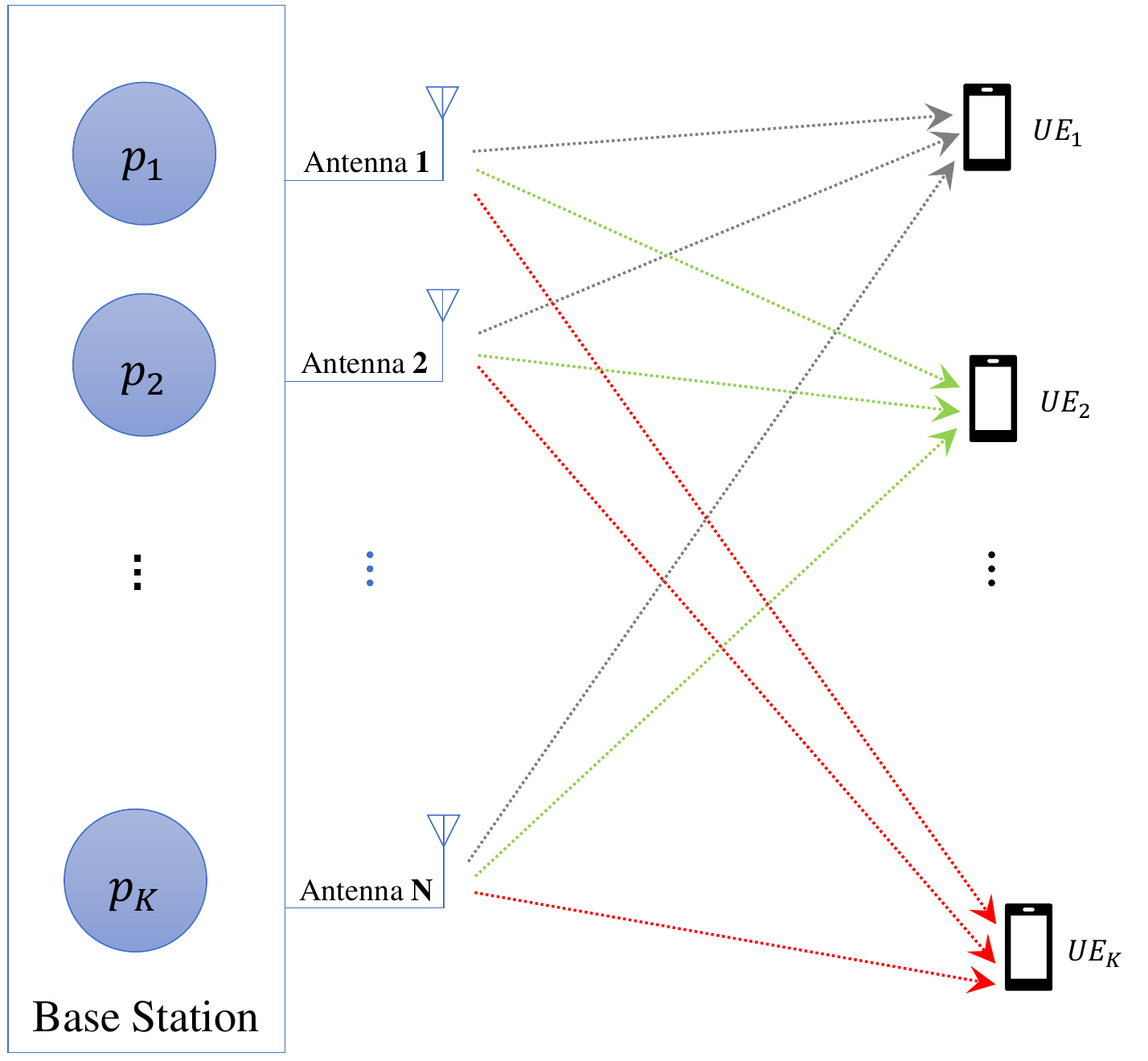}
\caption{Basic system model.} 
\label{fig:System-Model}
\end{figure}

We consider a downlink MU-MIMO wireless system model as shown in Fig. \ref{fig:System-Model}, where a transmitter (Base Station, BS) equipped with $N$ transmit antennas attempts to send $K$ equal size frames $\{p_1,p_2,...,p_K\}$ to single-antenna receivers (User Equipment, UE). The set of UE{s} is denoted as $k = \{1,2,...,K\}$.  

\section{Highly-Reliable XOR-assisted Transmission with new  Encoded OSTBC Approach}
\label{sec:High-Reliable XOR and OSTBC Approach}

\subsection{Transmitter Processing}
\label{subsec:Transmitter}

Following a similar side information (SI)-assisted error recovery technique considered in \cite{haghighat2017high}, the same network coding (NC) methodology used for a pair of users ($UE_1,UE_2$) is adapted for each two successive users in our proposed downlink MU-MIMO wireless system model. Similarly, the SI is disseminated through the use of superposition modulation to avoid the use of any extra physical resources and thus, the overall spectral efficiency is not decreased. 

\begin{figure}[t]
\centering
\includegraphics[width=0.9\columnwidth]{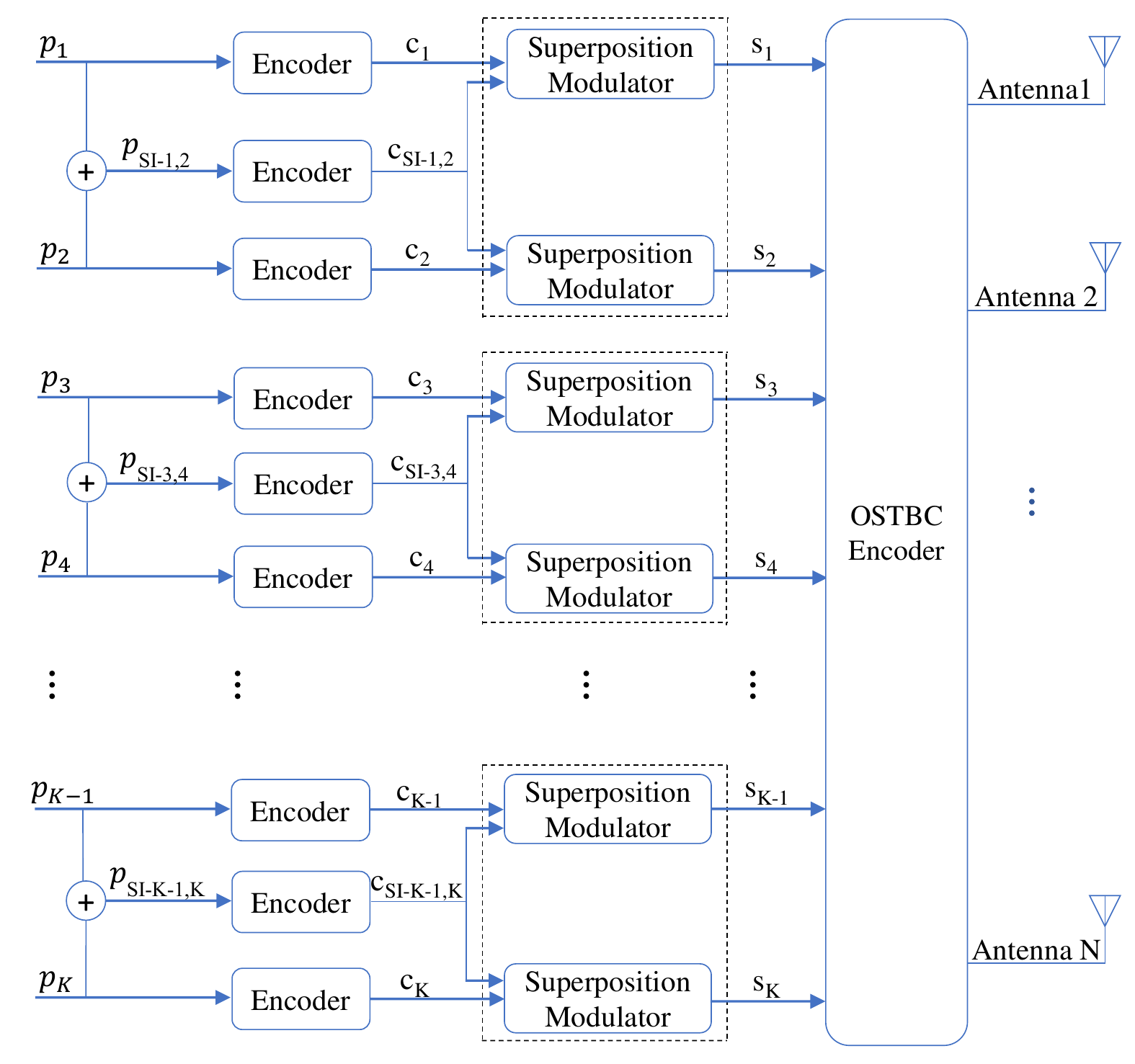}
\caption{Transmitter architecture.} 
\label{fig:Transmitter architecture}
\end{figure}

Fig. \ref{fig:Transmitter architecture} shows the basic transmitter processing for transmitting a number of $K$ equal size frames $\{p_1,p_2,...,p_K\}$ to a set of $K$ UE{s}. $K$, which represents the number of UE{s} supported by the system and equivalently the number of frames, is defined according to the new encoded OSTBC approach code which is basically related to the number of transmit antennas ($N$) at the BS. The number of UE{s} is strictly considered to be an even number. The restriction ensures the deployment of the side information frames $p_{SI-i,i+1}$, which are the bit-level XOR-ed of the two successive main user information frames $p_{i}$ and $p_{i+1}$, where $i$ is odd and $ i \in \{1,3,...,K-1\}$.

First, the frames $p_{i}$, $p_{i+1}$ and $p_{SI-i,i+1} = p_{i}\oplus p_{i+1}$ are encoded independently by rate-R convolutional encoder to produce the coded bit sequences $c_{i}$, $c_{i+1}$ and $c_{SI-i,i+1}$, where $i$ is odd and $i \in \{1,3,...,K-1\}$. Then, the generated code streams are fed into the symbol mapper to produce the transmit symbols for the $K$ set of users. Superposition modulation is employed in such a way that each three coded bit sequences $c_{i}$, $c_{i+1}$ and $c_{SI-i,i+1}$ are mapped to two composite constellation symbol sequences $s_{i}$ and $s_{i+1}$, $i$ is odd and $i \in \{1,3,...,K-1\}$ using only one physical resource $\mathcal{R}$ since a new OSTBC encoder approach is considered in the next processing stage.

\subsection{Superposition Modulation and Constellation Design}
\label{subsec:Superposition Modulation}

Following a similar procedure as gievn in\cite{haghighat2017high},\cite{nora}, a superposition constellation can be obtained by suitably combining two uniform-QAM constellations.

\begin{equation}
\label{eq:superposition-mod}
s_i = \sqrt{1-\alpha}x_1 + \sqrt{\alpha} x_2, \ i \in \{1,2,...,K\},
\end{equation}
where $x_1$ and $x_2$ are the constituent complex symbols, for example, m-QAM and the parameter ${0\leq } \alpha {\leq 1}$ is the power split between the two constellations.
\begin{figure}[t]
\centering
\includegraphics[width=0.9\columnwidth]{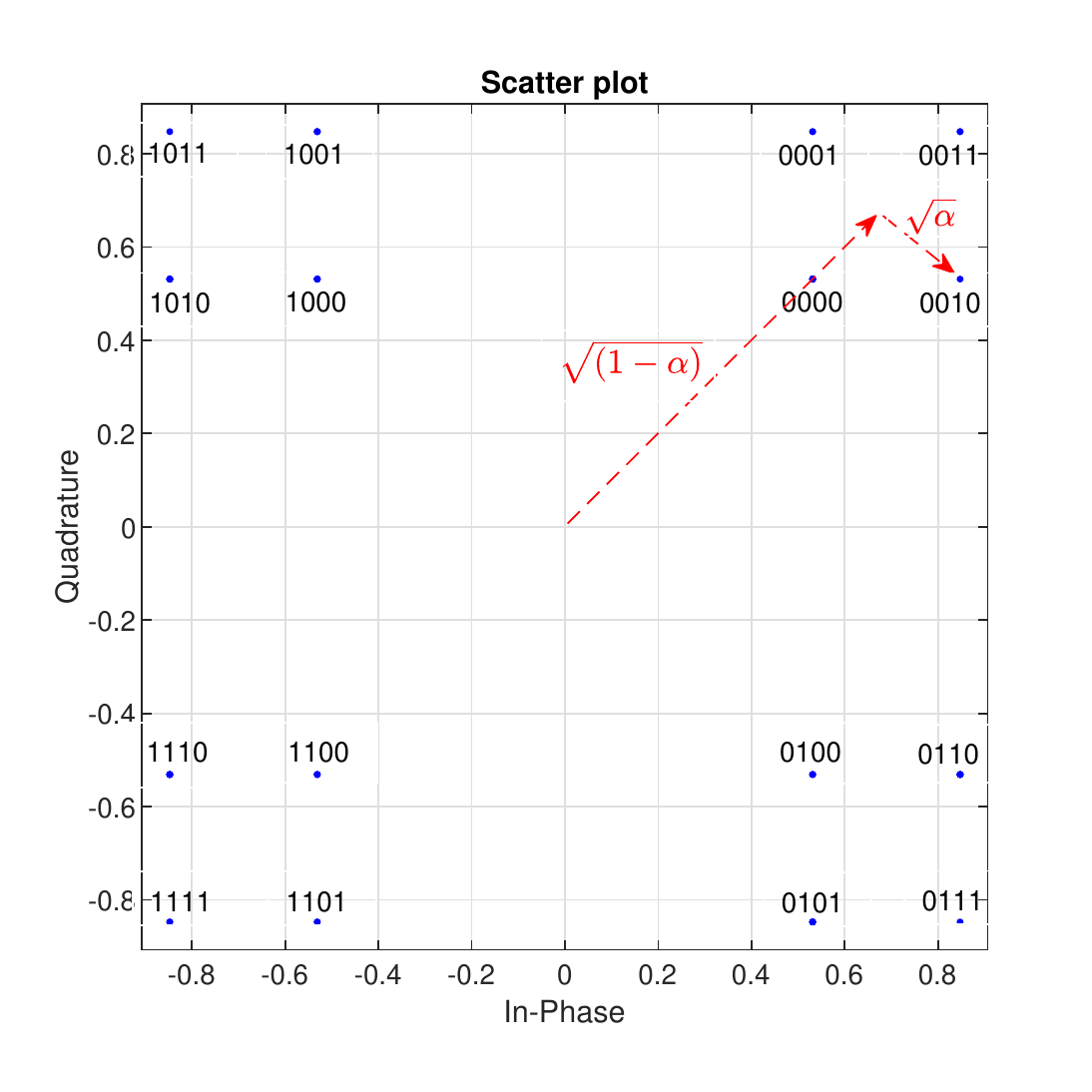}
\caption{Composite constellation with gray mapping $\alpha = 0.05$.} 
\label{fig:constellation}
\end{figure}
With the determination of $\alpha$ and the component (primary) modulations, the composite constellation is formed. Fig. \ref{fig:constellation} demonstrates a non-uniform composite constellation generated by the superimposition of two 4-QAM constellations with power split $\alpha = 0.05$. 
The superposition modulation is implemented such that, the number of $m_c^{(MSB)}$ bits from $c_{i}$ or $c_{i+1}$ is gathered with a number of $m_c^{(LSB)}$ bits from $c_{SI-i,i+1}$, $i$ is odd and $i \in \{1,3,...,K-1\}$ to build a complex symbol. The resulting composite constellation has a size of $M=2^{(m_c^{(MSB)}+m_c^{(LSB)})}$ where the most significant bits (MSBs) and the least significant bits (LSBs) of each symbol are represented by the $m_c^{(MSB)}$ and $m_c^{(LSB)}$, respectively. For our proposed system model two lower order 4-QAM constellations are adopted such that $m_c^{(MSB)}=m_c^{(LSB)}=2$, resulting in 16 points composite constellations. In this case, the quadrant of the transmitted symbol is specified by the bits from $c_{i}$ or $c_{i+1}$, while the bits from $c_{SI-i,i+1}$ decide the correct area of the transmitted symbol in the quadrant.

\subsection{MU-MIMO New Encoded OSTBC Approach}
\label{subsec:MU-MIMO OSTBC Approach}
A systematic design method to generate high rate complex orthogonal space-time block codes (STBCs) for any number of transmit antennas as presented in \cite{su2004systematic} is used for our system model, however, with a new approach.

The proposed STBC is described with a  $P$x$N$ transmission matrix $G_N$, the entries of the matrix $G_N$ are linear combinations of the precise users symbols $\{s_1,s_2,...,s_K\}$ and their conjugates, where $P$ refers to the block length of the matrix, $N$ and $K$ represent  the number of transmit antennas and number of users, respectively.
In this work, we consider $G_2$ and $G_4$ which represent codes that utilize two and four transmitting antennas, respectively. According to the systematic design algorithm to generate high rate complex orthogonal STBCs for any number of transmit  antennas, $G_2$ and $G_4$ are given by
\begin{equation}
\label{eq:G2 matrix}
G_2 = 
\begin{bmatrix}
	s_1  & s_2 \\ -s^*_2 & s^*_1
	\end{bmatrix},
\end{equation}

\begin{equation}
\label{eq:G4 matrix}
G_4 = 
\begin{bmatrix}
	   s_1  & s_2 & s_3  & 0
	\\ -s^*_2 & s^*_1 & 0 & s^*_4
	\\ -s^*_3 & 0 & s^*_1 & s^*_5
	\\ 0 & -s^*_3 & s^*_2 & s^*_6
	\\ 0 & -s_4 & -s_5 & s_1
	\\ s_4 & 0 & -s_6 & s_2
	\\ s_5 & s_6 & 0 & s_3
	\\ -s^*_6 & s^*_5 & -s^*_4 & 0
	\end{bmatrix}.
\end{equation}

 As stated earlier, our transmission model uses a superposition modulation with a total of $M = 2^b $ composite constellation elements. In one time slot, $b$ specific user bits reach the encoder and the appropriate constellation signal corresponding to that user is picked, by setting $s_i$ as the constellation symbol selected for specific bits of user $k$, $i=k$ and $i \in \{1,2,...,K\}$, a matrix $G_N$ is realized with entries as linear combinations of $\{s_1,s_2,...,s_K\}$ and their conjugates. Here $K$ is the number of users that can be supported by the MU-MIMO system for a chosen number of transmit antennas $N$, which can be defined also by how many different entries $s_i$  are generated in the matrix $G_N$. For $G_2$ and $G_4$, the proposed system accommodates $K=2$ and $K=6$ users, respectively.
 
 As $P$ time slots are used to transmit one symbol for each user, the rate of the code is $1/P$, resulting in $1/2$ and $1/8$ code rates for $G_2$ and $G_4$, respectively. One can notice that the rates are low when compared to the use of conventional OSTBC, where the rates can be defined as $1$ and $3/4$ for $G_2$ and $G_4$, respectively. 
 Basically, the orthogonality of the columns of $G_N$ for this specific MU-MIMO OSTBC approach allows the application of a simple linear decoding scheme at the receiver and also the use of only one common physical resource $\mathcal{R}$  for all users.
 
 \subsection{Decoding Algorithm and Sub-Constellation Alignment for Signal Combining}
\label{subsec:Decoding Algorithm and Sub-Constellation Alignment for Signal Combining}

\begin{figure}[t]
\centering
\includegraphics[width=0.9\columnwidth]{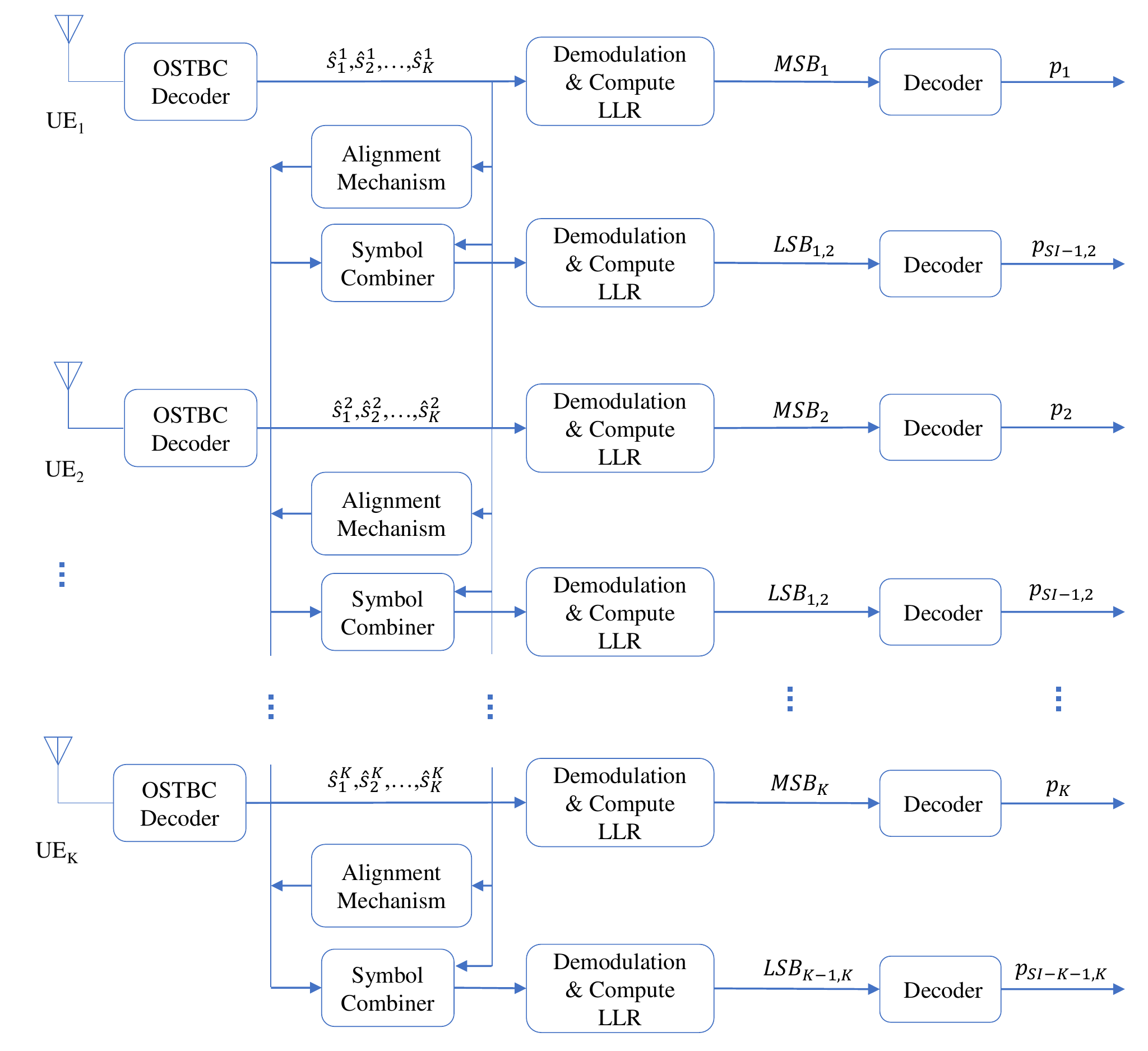}
\caption{Receiver architecture.} 
\label{fig:Receiver architecture}
\end{figure}

A sub-constellation alignment approach is presented and investigated in\cite{haghighat2017high} for joining of LSBs of the received symbols to guarantee proper delivery of SI. The sub-constellation alignment for signal combining procedure is modified for the new MU-MIMO encoded OSTBC system model considered in this work. Fig. \ref{fig:Receiver architecture} demonstrates the general construction of the receiver.

As shown earlier, for $N = 2$ transmitting antennas, a $G_2$ matrix is generated as in (\ref{eq:G2 matrix}). In this case, $K = 2$ single-antenna UEs are served, the received signals are given by,
\begin{equation}
\label{eq:received signals}
\begin{cases} y^1_k = h_{1,k}s_1 + h_{2,k}s_2 + n^1_k, & \mbox{at time one,} \\ y^2_k = - h_{1,k}s^*_2 + h_{2,k}s^*_1 + n^2_k, & \mbox{at time two, $k \in \{1,2\}$}. \end{cases}
\end{equation}
The coefficient $h_{1,k}$ is the channel from transmit antenna one to receive antenna of user $k$ and $h_{2,k}$ is the channel from transmit antenna two to receive antenna  of user $k$. $ n^1_k$  and $n^2_k$  are AWGN of receiver $k$ with zero mean and variance $N_0$ at time one and two, respectively. 
The OSTBC combining scheme builds the following two signals at the level of each user $k$.
\begin{equation}
\label{eq:OSTBC combined signals}
\begin{cases} \hat{s}^k_1 = y^1_k h^*_{1,k} + y^{2*}_k h_{2,k}, &  \\ \hat{s}^k_2 = y^1_k h^*_{2,k} - y^{2*}_k h_{1,k}, & \mbox{$k \in \{1,2\}$}. \end{cases}
\end{equation}

Similarly, the combining scheme for the MU-MIMO new encoded OSTBC approach can be generalized for any number of transmitting antennas $N$, in this case, $K$ users are considered. Thus according to the OSTBC combining scheme signals $\{\hat{s}^k_1,\hat{s}^k_2,...,\hat{s}^k_K\}$ are built at the level of each receiver $k$, $k \in \{1,2,...,K\}$.

As expressed before, the data corresponding to frames $\{p_1,p_2,...,p_K\}$ is transmitted utilizing MSBs of the transmitted composite constellation symbols. In this manner, it is less likely to make a mistake in decoding the main information frames $\{p_1,p_2,...,p_K\}$ than $p_{SI-i,i+1}$, $i$ is odd and $i \in \{1,3,...,K-1\}$ which depend on a transmission with smaller Euclidean distance. Nevertheless, since the equivalent $c_{SI}$ is transmitted within the composite symbols of every two successive users $(k,k+1)$, $k$ is odd and $k \in \{1,3,...,K-1\}$, These can be joined to compensate the loss for being sent with a smaller euclidean distance. 

After building $\{\hat{s}^k_1,\hat{s}^k_2,...,\hat{s}^k_K\}$ signals at the level of each user $k$, our main issue is the combining technique. However, since the MSBs of the successive transmitted composite symbols $s_{i}$ and $s_{i+1}$, $i$ is odd and  $i \in \{1,3,...,K-1\}$ are not the same, an immediate combining is not conceivable. Thu to address this issue, before the combining procedure, an arrangement or phase alignment should be performed.
The alignment mechanism should be performed on the OSTBC combined symbols $\{\hat{s}^k_1,\hat{s}^k_2,...,\hat{s}^k_K\}$ built at the level of each receiver $k$ as

\begin{equation}
\label{eq:alignment mechanism}
\bar{s}^k_i =\begin{cases}  Sgn(Re(\hat{s}^k_i)) \prod\limits^{K}_{j=1, j\neq{k}}Sgn(Re(\hat{s}^j_{i+1}))Re(\hat{s}^k_i) + \Im\\ Sgn(Im(\hat{s}^k_i))\prod\limits^{K}_{j=1, j\neq k}Sgn(Im(\hat{s}^j_{i+1}))Im(\hat{s}^k_i), \\\mbox{for $i$ odd,} \\ Sgn(Re(\hat{s}^k_i)) \prod\limits^{K}_{j=1, j\neq{k}}Sgn(Re(\hat{s}^j_{i-1}))Re(\hat{s}^k_i) + \Im\\ Sgn(Im(\hat{s}^k_i))\prod\limits^{K}_{j=1,j\neq k }Sgn(Im(\hat{s}^j_{i-1}))Im(\hat{s}^k_i),\\\mbox{for $i$ even, } \end{cases}
\end{equation}
where $i \in \{1,2,...,K\}$ and $k \in \{1,2,...,K\}$. $\Im = \sqrt{-1}$, $Sgn(x)$, $Re(x)$ and $Im(x)$ represent sign, real and imaginary functions, respectively.\par
After the sub-constellation alignment, the combined signals ${s}^k_{i,Comb}$ can be computed as,
\begin{equation}
\label{eq:combined signals}
{s}^k_{i,Comb} =\begin{cases}  \hat{s}^k_i + \sum\limits_{m=1,m \neq k}^{K} \bar{s}^m_{i+1}, \mbox{for $i$ odd.}\\
\hat{s}^k_i + \sum\limits_{m=1,m \neq k}^{K} \bar{s}^m_{i-1}, \mbox{for $i$ even. } \end{cases}
\end{equation}
where $i \in \{1,2,...,K\}$ and $k \in \{1,2,...,K\}$.\par
We provide the decoding formula for each symbol for the general case of $N$ transmit antennas and $K$ users where the maximum likelihood detection has to minimizing the decision statistic,
\begin{equation}
\label{eq:maximum likelihood decision}
|s^k_{i,Comb} - s_{j}|^2+ (-1 + \sum\limits_{m=1}^{K} \sum\limits_{n=1}^{N} |h_{n,m}|^2) |s_{j}|^2 
\end{equation}
over all possible values of  constellation symbols $s_{j}$ for decoding $s^k_{i,Comb}$, where $h_{n,m}$ is the channel from transmit antenna $n$ to receive antenna of user $m$, $i \in \{1,2,...,K\}$ and $k \in \{1,2,...,K\}$.

\subsection{SI-assisted Decoding Strategy}
\label{subsec:Decoding Strategy}
As a consequence of the new MU-MIMO encoded OSTBC approach, $UE_k$ not only receives its own symbol but all other users transmitted symbols, the offered diversity allows the recovery of a specific user frame in different ways. If $K$ users are considered, for each $UE_k$, $k \in \{1,2,...,K\}$, we provide $2K$ different ways to recover the user specific frame $p_i$. As an example, for $N=2$ transmit antennas, this implies $K=2$ users in the downlink. First, $UE_1$ attempts to decode $c_1$ and in case of a decoding error, $UE_1$ sets out on decoding $c_2$ and $c_{SI-1,2}$. On the off chance that decoding of both $c_2$ and $c_{SI-1,2}$ is effective, $UE_1$ obtains $p_1$ from successfully decoded $p_2$ and $p_{SI-1,2}$. However, if there is an error in the decoding of both $c_2$ and $c_{SI-1,2}$, $UE_1$ has two additional chances to decode $p_1$ since $UE_2$ receives a second version of the transmitted frames. Now, the focus is to decode $c_1$ received by the second user. In the event of a decoding error, the last option for the first user is to decode $c_2$ and $c_{SI-1,2}$, with the condition that decoding of both is successful, $UE_1$ recovers $p_1$, however, if either of the two is decoded erroneously, $UE_1$ considers the frame $p_1$ as lost.\par
Let us consider the same example, where the BS is equipped with two transmit antennas and two users are considered. Let $P^{(i)}_{e,k}$, $k \in \{1,2\}$ and $i \in \{1,2,SI-1,2\}$, indicates the probability of error in decoding frame $p_i$ of user $k$. $P^{(i)}_{c,NC}$ signifies the probability of correct recovery of $p_i$ and can be just expressed as
\begin{multline}
\label{eq:probability of correct}
P^{(i)}_{c,NC} = 1 -P^{(i)}_{e,1} + P^{(i)}_{e,1}\big(1-P^{(j)}_{e,1}\big)\big(1-P^{(SI-1,2)}_{e,1}\big) \\
  +P^{(i)}_{e,1}\big(1-P^{(j)}_{e,1}\big)\big(1-P^{(SI-1,2)}_{e,1}\big)P^{(i)}_{e,2} 
  \\+P^{(i)}_{e,1} \big(1-P^{(j)}_{e,1} \big) \big(1-P^{(SI-1,2)}_{e,1}\big)P^{(i)}_{e,2}\big(1-P^{(j)}_{e,2}\big)\big(1-P^{(SI-1,2)}_{e,2}\big),
\end{multline}
for $i,j \in \{1,2\}, i \neq j$. 
In \eqref{eq:probability of correct}, $P^{(i)}_{e,k}$ and $P^{(j)}_{e,k}$ are assumed to be mutually exclusive from $P^{(SI-1,2)}_{e,k}$. Consequently the probability of error of $p_i$ is denoted as $P^{(i)}_{e,NC}$, which is $P^{(i)}_{e,NC} = 1 - P^{(i)}_{c,NC}$, is given by  
\begin{multline}
\label{eq:probability of error}
P^{(i)}_{e,NC} = P^{(i)}_{e,1} \big[1 - \big(1-P^{(j)}_{e,1}\big)\big(1-P^{(SI-1,2)}_{e,1}\big) \\
  - \big(1-P^{(j)}_{e,1}\big)\big(1-P^{(SI-1,2)}_{e,1}\big)P^{(i)}_{e,2}\\ 
  - \big(1-P^{(j)}_{e,1} \big) \big(1-P^{(SI-1,2)}_{e,1}\big)P^{(i)}_{e,2}\big(1-P^{(j)}_{e,2}\big)\big(1-P^{(SI-1,2)}_{e,2}\big)\big]. 
\end{multline}
In general, this can be seen as a Markov chain with memory, or a Markov chain of order $2K$. As stated previously, $2K$ represents the number of chances provided to recover the user specific frame. Thus, equations \eqref{eq:probability of correct} and \eqref{eq:probability of error} can be rewritten in the same manner with the possibility to consider the general model with $N$ transmit antennas and $K$ users.

\section{Simulation Results}
\label{sec:results}
In the following, we provide simulation results in order to validate the error performance improvement for two different MU-MIMO transmission scenarios. In the first scenario we consider two transmit antennas at the base station, $G_2$ is considered for the new encoded OSTBC approach with rate $1/2$ and two users can be accommodated by the system. Whereas in the second scenario, four transmit antennas are utilized at the base station, which involves the use of $G_4$ for the new encoded OSTBC approach with rate $1/8$, and six users are considered. We assume that the same amount of energy is shared by the transmitting antennas to guarantee that the same aggregate power is utilized in both scenarios, where the symbol energy ${E_s} = \frac{1}{2}$ and  ${E_s} = \frac{1}{4}$ are the average energies of the composite constellation for the first and second scenarios, separately.

A rate-1/3 convolutional encoder, $[133, 171, 165]_8$ with constraint length 6 is employed along with superposition modulation with the choice of $m_c^{(MSB)}=m_c^{(LSB)}=2$ which results in a $16$ point gray mapping composite constellation that is composed of two lower order 4-QAM constellations, the power parameter $\alpha$ is chosen to be $0.05$ and $0.10$. The block length is set to $100$ bits and the channels involved are modeled by independent Rayleigh fading $\mathcal{CN}(0,1)$. The soft-decision Viterbi decoding is carried out at the receiver side.

In all the figures provided in this section, the legend entries $P_{i,NC}$ and $P_{i}$ refer to the error performance of frame $p_{i}$ of any user $k$, where $i \in \{1,2,...,K\}$ and similarly $k \in \{1,2,...,K\}$, with and without the SI-assisted decoding strategy (detailed for the case of first scenario in subsection \ref{subsec:Decoding Strategy}), respectively. For the evaluation of ``LSB signal combined" the performance of $P_{SI}$  for any arbitrary $p_{SI-i,i+1}$, $i$ is odd and $i \in \{1,3,...,K-1\}$, is considered as well, while the legend section ``LowerBound'' illustrates the anticipated lower bound (as given by \eqref{eq:probability of error} for the case of the first scenario) as a reference for comparison.

\begin{figure}[t]
\centering
\includegraphics[width=0.9\columnwidth]{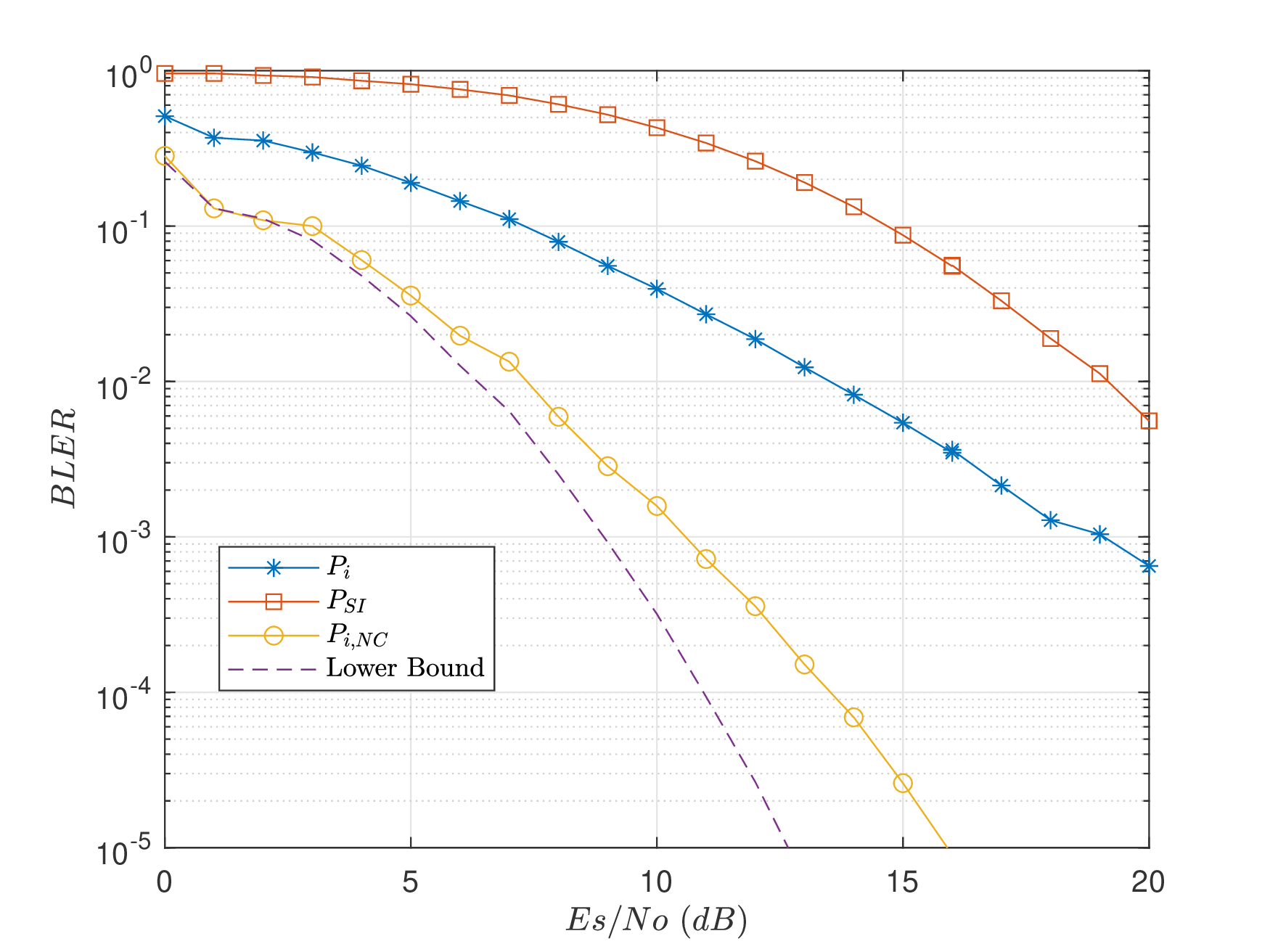}
\caption{BLER performance for the case of two transmit antennas  with new encoded OSTBC approach and $\alpha=0.05$.} 
\label{fig:2TX_alpha0.05}
\end{figure}

\begin{figure}[t]
\centering
\includegraphics[width=0.9\columnwidth]{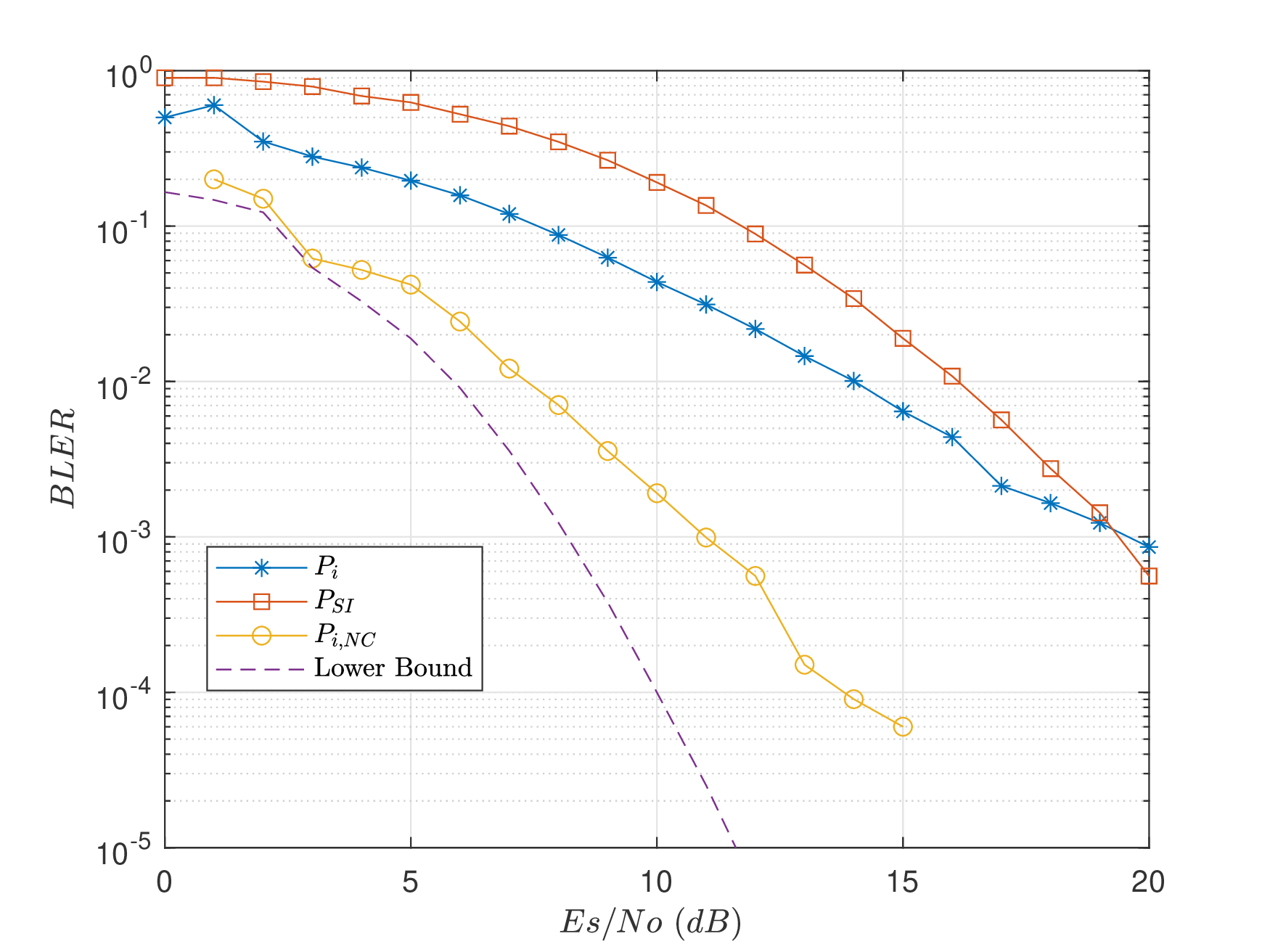}
\caption{BLER performance for the case of two transmit antennas  with new encoded OSTBC approach and $\alpha=0.10$.}
\label{fig:2TX_alpha0.1}
\end{figure}

\begin{figure}[t]
\centering
\includegraphics[width=0.9\columnwidth]{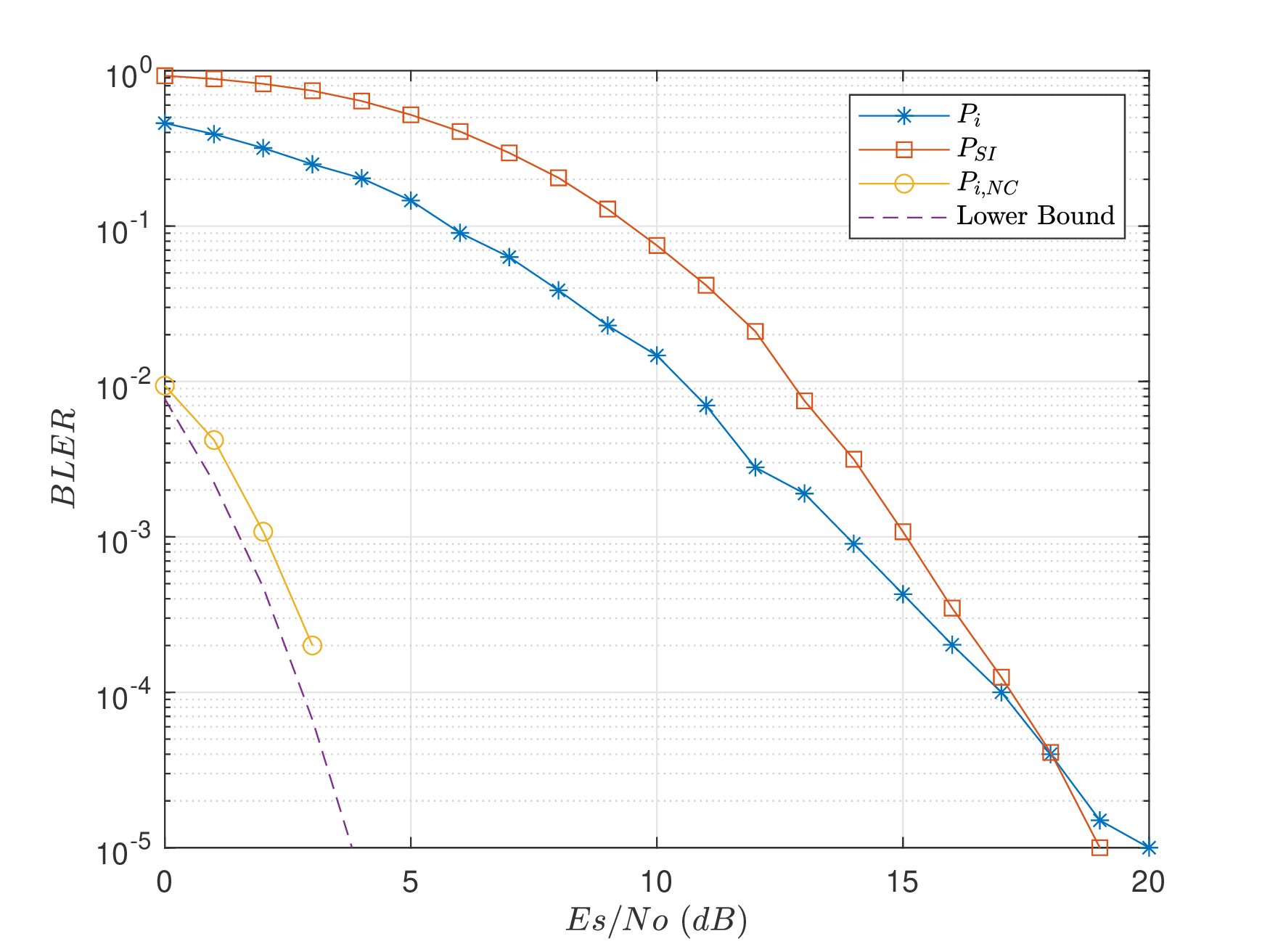}
\caption{BLER performance for the case of four transmit antennas with new encoded OSTBC approach and $\alpha=0.05$.} 
\label{fig:4TX_alpha0.05}
\end{figure}

\begin{figure}[t]
\centering
\includegraphics[width=0.9\columnwidth]{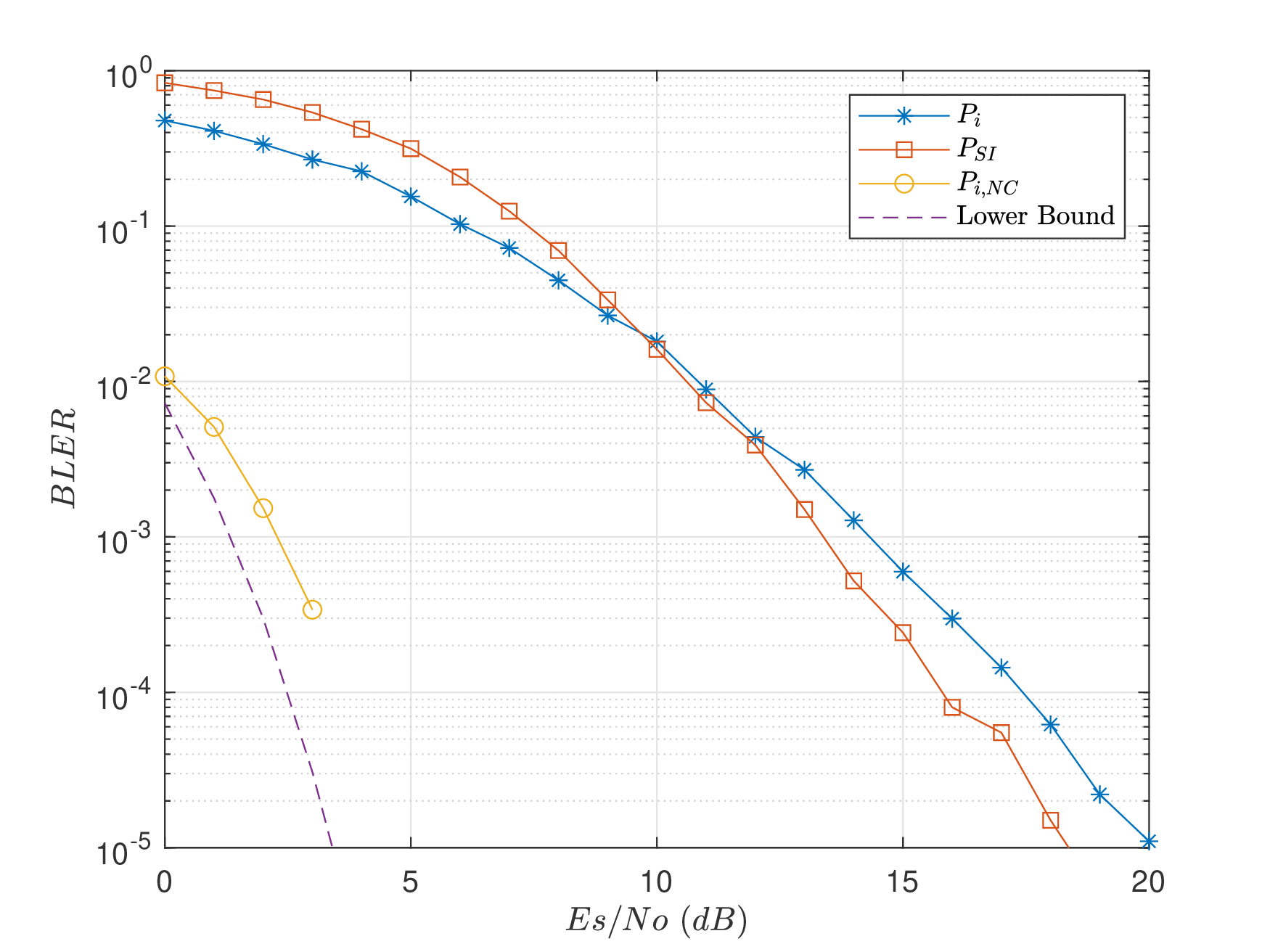}
\caption{BLER performance for the case of four transmit antennas with new encoded OSTBC approach and $\alpha=0.10$.}
\label{fig:4TX_alpha0.1}
\end{figure}

Fig. \ref{fig:2TX_alpha0.05} and Fig. \ref{fig:2TX_alpha0.1} demonstrate the block error rate (BLER) performance over $\frac{E_s}{N_0}$ corresponding to the first scenario ($N=2$ and $K=2$) for two different estimations of $\alpha = 0.05$ and $\alpha = 0.10$, respectively. To verify the performance gains promised by the proposed new encoded OSTBC approach for a MU-MIMO scenario and appropriate combining of received signals for error recovery, one can refer to \cite{haghighat2017high}; hence a significant achieved performance gain can be clearly seen. This is due to the additional diversity offered by the new encoded OSTBC approach, where each user is receiving all other users symbols. On one hand, this fact will enhance the alignment mechanism for signal combining which now exploits the all other users channels and, on the other hand, several chances for the recovery of frame $p_i$ of user $k$ are offered, $2K$ ($= 4$) alternative ways; hence reliability is strongly supported. 

Fig. \ref{fig:2TX_alpha0.1} shows the impact of of picking $\alpha$, the figure affirms the significant $P_{SI}$ gain by the proposed signal combining, that even outperforms the performance of $P_{i}$ at higher $E_s/N_0$ for $\alpha=0.10$. However, $\alpha$ should be picked in such a way to help a moderately comparable performance for $p_{i}$ and $p_{SI-i,i+1}$ transmission. A poorly chosen value of $\alpha$ would sway the power to one of the constitute constellations that leads to a degradation of the overall performance.

Fig. \ref{fig:4TX_alpha0.05} and Fig. \ref{fig:4TX_alpha0.1} show the block error rate (BLER) performance over $\frac{E_s}{N_0}$ corresponding to the second scenario ($N=4$ and $K=6$) for two different values of $\alpha = 0.05$ and $\alpha=0.10$, respectively. We observe that the performance is significantly improved when compared to the first scenario. Despite the fact that the rate of the new encoded OSTBC code is brought down to $1/8$, more users are supported by the system which strengthens the LSB signal combining and provides $2K (=12)$ alternative attempts for user frame recovery.

As a global evaluation for the proposed system, multiple users can be accommodated by sharing just a single resource; furthermore, as more users are added, a higher reliability is promised. Apart from that, one can consider the fact of lowering the transmission rate where latency comes into account. To compensate for the low rates, simulation is conducted for both scenarios where conventional OSTBC is used with rates $1$ and $3/4$ for G2 and G4, respectively. As a penalty to pay, $K$ orthogonal resources are needed for $K$ users.
\begin{figure}[t]
\centering
\includegraphics[width=0.9\columnwidth]{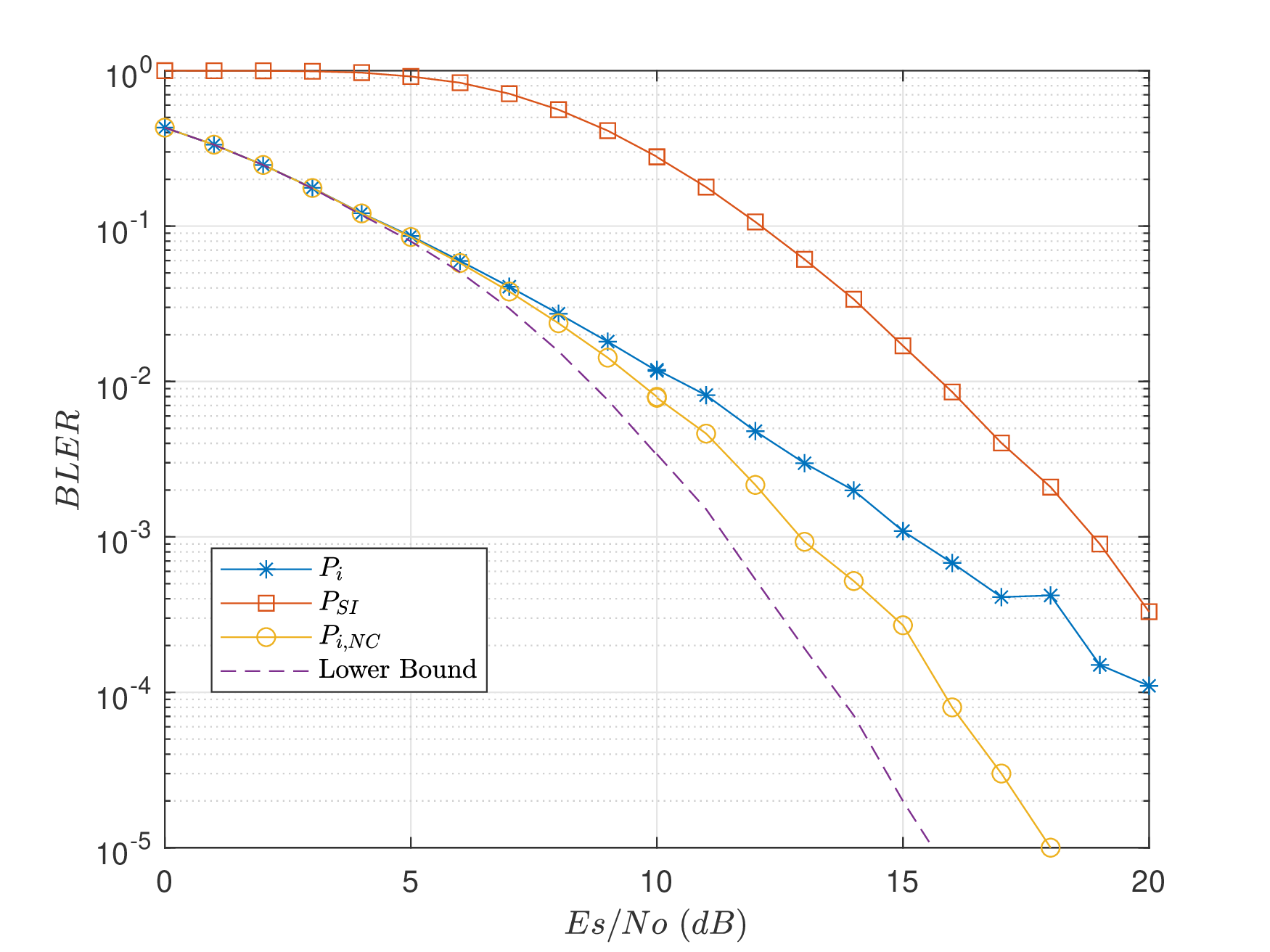}
\caption{BLER performance for the case of two transmitting antennas with conventional encoded OSTBC and $\alpha=0.05$.} 
\label{fig:2TX_convSTBC_alpha0.05}
\end{figure}

\begin{figure}[t]
\centering
\includegraphics[width=0.9\columnwidth]{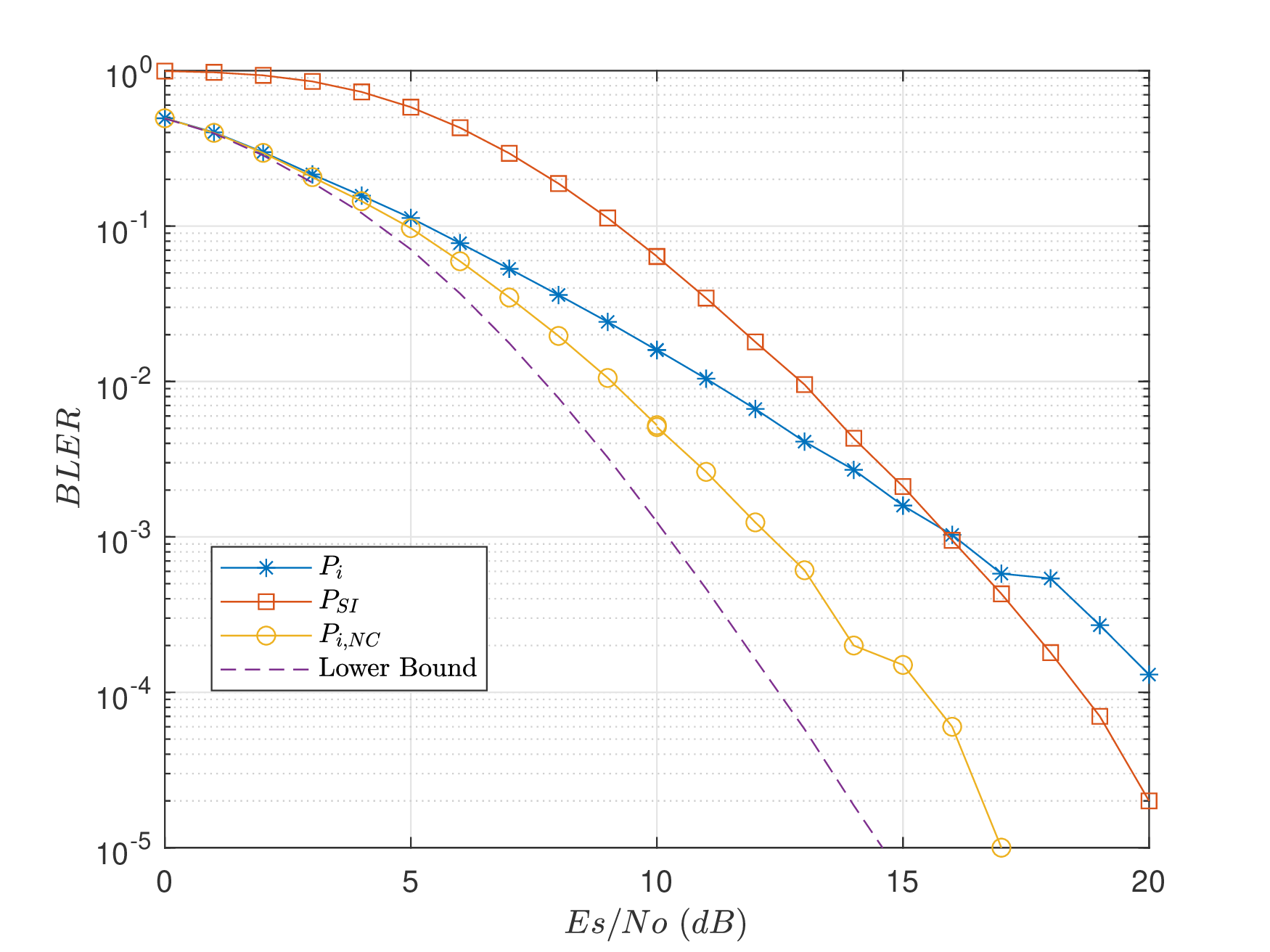}
\caption{BLER performance for the case of two transmitting antennas with conventional encoded OSTBC and $\alpha=0.10$.}
\label{fig:2TX_convSTBC_alpha0.1}
\end{figure}

Fig. \ref{fig:2TX_convSTBC_alpha0.05} and Fig. \ref{fig:2TX_convSTBC_alpha0.1} present the block error rate (BLER) performance over $\frac{E_s}{N_0}$ for the situation where $N=2$, $K=2$ and conventional OSTBC is used for two different values of $\alpha = 0.05$ and $\alpha=0.10$, respectively. Comparing Fig. \ref{fig:2TX_convSTBC_alpha0.05} and Fig. \ref{fig:2TX_convSTBC_alpha0.1} with Fig. \ref{fig:2TX_alpha0.05} and Fig. \ref{fig:2TX_alpha0.1}, we notice the performance gain accomplished by $P_i$ and $P_{SI}$ when conventional OSTBC is used. In Fig. \ref{fig:2TX_convSTBC_alpha0.05} and Fig. \ref{fig:2TX_convSTBC_alpha0.1}, the performance effect of $P_{i,NC}$ is apparent at a considerably higher $E_s/N_0$ range. However, the performance achieved is less compared with results shown in Fig. \ref{fig:2TX_alpha0.05} and Fig. \ref{fig:2TX_alpha0.1} where the new encoded OSTBC approach is considered. This can be justified by the fact that the diversity offered by the new encoded OSTBC (a user not only receives its own symbol but all other users transmitted symbols) no longer exists, by utilizing the classical OSTBC every user is just receiving its own symbol. Thus, the alignment mechanism and symbol combining is valid only between two successive users, or in other words, the alternative attempts given by the SI-assisted decoding strategy do not rely upon the number of users K, only two different ways for frame recovery are possible regardless of the number of users in the system.

\begin{figure}[h]
\centering
\includegraphics[width=0.9\columnwidth]{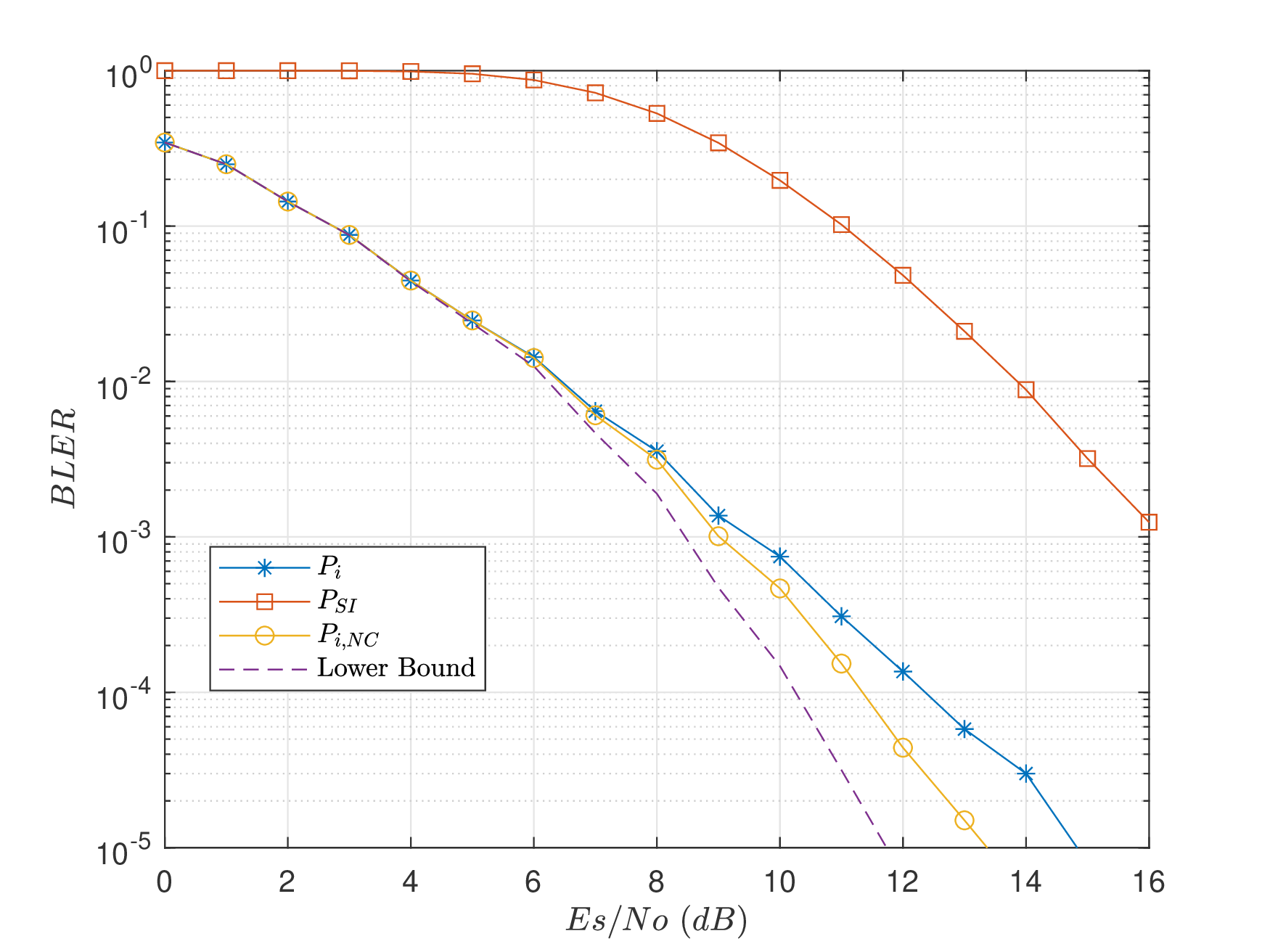}
\caption{BLER performance for the case of four transmitting antennas with conventional encoded OSTBC and $\alpha=0.05$.} 
\label{fig:4TX_convSTBC_alpha0.05}
\end{figure}
\begin{figure}[h]
\centering
\includegraphics[width=0.9\columnwidth]{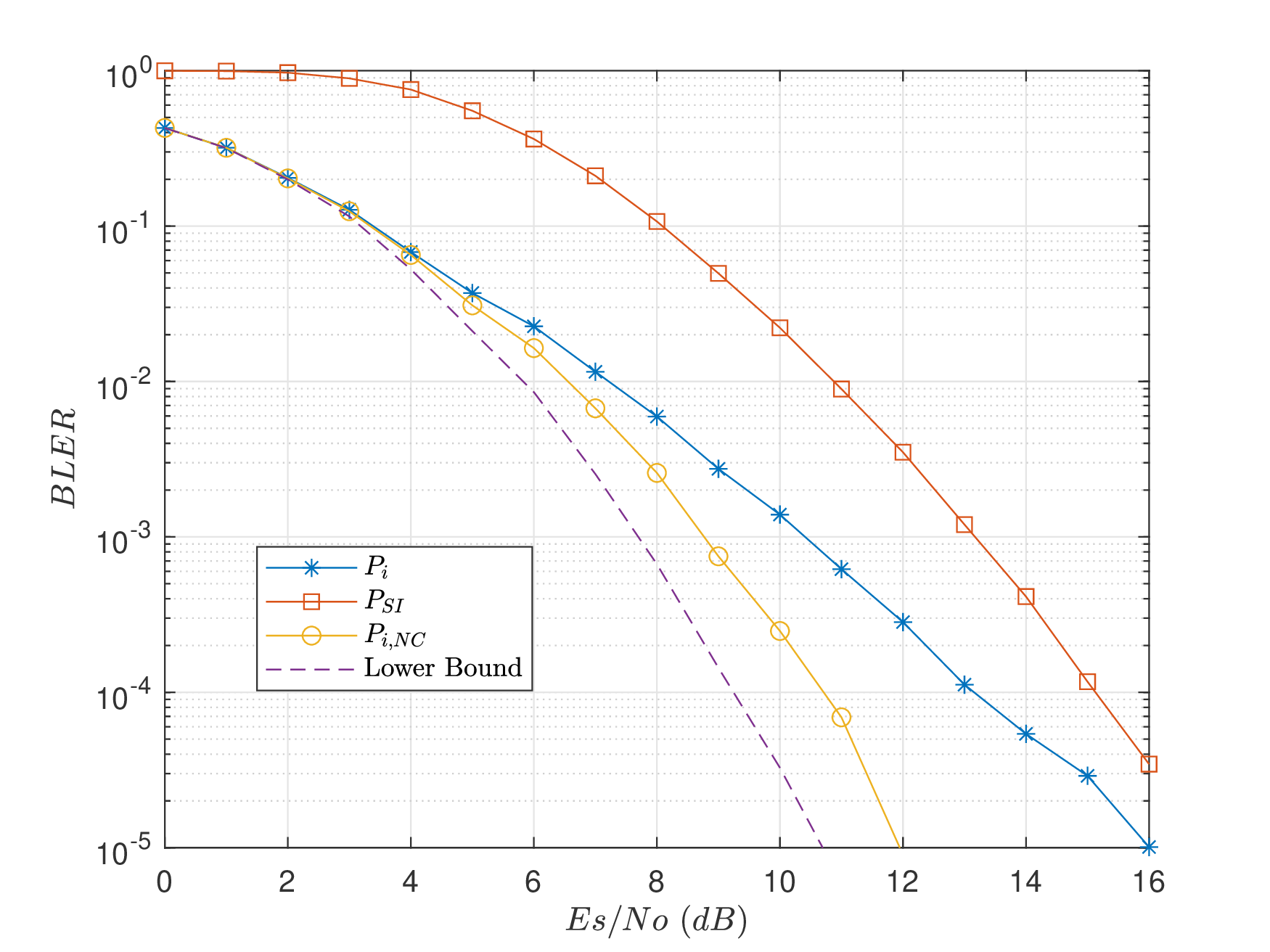}
\caption{BLER performance for the case of four transmitting antennas with conventional encoded OSTBC and $\alpha=0.10$.}
\label{fig:4TX_convSTBC_alpha0.1}
\end{figure}
Fig. \ref{fig:4TX_convSTBC_alpha0.05} and Fig. \ref{fig:4TX_convSTBC_alpha0.1} illustrate the block error rate (BLER) performance over $\frac{E_s}{N_0}$ for the situation where $N=4$, $K=6$ and conventional OSTBC is used for two different values of $\alpha = 0.05$ and $\alpha=0.10$, respectively. The results provided demonstrate that significant gains can be achieved compared with results shown in Fig. \ref{fig:2TX_convSTBC_alpha0.05} and Fig. \ref{fig:2TX_convSTBC_alpha0.1}. This is mainly due to the conventional OSTBC when increasing the number of transmit antennas.

\section{Conclusion}
\label{sec:Conclusion}
In this work a downlink MU-MIMO scheme to achieve highly reliable transmission is investigated with the use of  a new encoded OSTBC approach and superposition modulation. Furthermore, a constructive combining of LSB's of the received symbols through sub-constellation alignment is employed. The block error rate (BLER) simulation results show the significant performance gain obtained by the proposed system.  Enhanced reliability is seen with an increased number of users at the expense of lowering the rate. Another advantage offered by the new encoded OSTBC approach is that multiple users can be accommodated by sharing only a single resource. Due to the higher expected latency with the number of users, simulation results are also shown when conventional OSTBC is utilized where high rate codes are incorporated. A significantly higher reliability is realized at the expense of employing as much as orthogonal resources as the number of users considered. Simulation results also show that the reliability can be further improved in this case by increasing the number of transmit antennas at the base station.


\vspace{-1px}
\bibliographystyle{IEEEtran}
\bibliography{IEEEabrv,ref_conf_short,ref_jour_short,references}

\begin{thebibliography}{10}
\providecommand{\url}[1]{#1}
\csname url@samestyle\endcsname
\providecommand{\newblock}{\relax}
\providecommand{\bibinfo}[2]{#2}
\providecommand{\BIBentrySTDinterwordspacing}{\spaceskip=0pt\relax}
\providecommand{\BIBentryALTinterwordstretchfactor}{4}
\providecommand{\BIBentryALTinterwordspacing}{\spaceskip=\fontdimen2\font plus
\BIBentryALTinterwordstretchfactor\fontdimen3\font minus
  \fontdimen4\font\relax}
\providecommand{\BIBforeignlanguage}[2]{{%
\expandafter\ifx\csname l@#1\endcsname\relax
\typeout{** WARNING: IEEEtran.bst: No hyphenation pattern has been}%
\typeout{** loaded for the language `#1'. Using the pattern for}%
\typeout{** the default language instead.}%
\else
\language=\csname l@#1\endcsname
\fi
#2}}
\providecommand{\BIBdecl}{\relax}
\BIBdecl

\bibitem{Mueck-5GCHAMPION-2016}
{M. Mueck, et al.}, ``{5G CHAMPION - Rolling out 5G in 2018},'' in \emph{Proc.
  IEEE Global Commun. Conf. Workshops}, Dec. 2016, pp. 1--6.

\bibitem{johansson2015radio}
N.~A. Johansson, Y.-P.~E. Wang, E.~Eriksson, and M.~Hessler, ``Radio access for
  ultra-reliable and low-latency 5g communications,'' in \emph{Communication
  Workshop (ICCW), 2015 IEEE International Conference on}.\hskip 1em plus 0.5em
  minus 0.4em\relax IEEE, 2015, pp. 1184--1189.

\bibitem{nazer2011reliable}
B.~Nazer and M.~Gastpar, ``Reliable physical layer network coding,''
  \emph{Proceedings of the IEEE}, vol.~99, no.~3, pp. 438--460, 2011.

\bibitem{haghighat2017high}
A.~Haghighat and S.~P. Herath, ``High reliability downlink transmission with
  superposition modulated side information,'' in \emph{Wireless Communications
  and Networking Conference (WCNC), 2017 IEEE}.\hskip 1em plus 0.5em minus
  0.4em\relax IEEE, 2017, pp. 1--6.

\bibitem{hoeher2011superposition}
P.~A. Hoeher and T.~Wo, ``Superposition modulation: myths and facts,''
  \emph{IEEE Communications Magazine}, vol.~49, no.~12, 2011.

\bibitem{alamouti1998simple}
S.~M. Alamouti, ``A simple transmit diversity technique for wireless
  communications,'' \emph{IEEE Journal on selected areas in communications},
  vol.~16, no.~8, pp. 1451--1458, 1998.

\bibitem{tarokh1998application}
V.~Tarokh, H.~Jafarkhani, and A.~Calderbank, ``The application of orthogonal
  designs to wireless communication,'' in \emph{Information Theory Workshop,
  1998}.\hskip 1em plus 0.5em minus 0.4em\relax IEEE, 1998, pp. 46--47.

\bibitem{tarokh1999space}
V.~Tarokh, H.~Jafarkhani, and A.~R. Calderbank, ``Space-time block codes from
  orthogonal designs,'' \emph{IEEE Transactions on Information theory},
  vol.~45, no.~5, pp. 1456--1467, 1999.

\bibitem{su2004systematic}
W.~Su, X.-G. Xia, and K.~R. Liu, ``A systematic design of high-rate complex
  orthogonal space-time block codes,'' \emph{IEEE Communications Letters},
  vol.~8, no.~6, pp. 380--382, 2004.

\bibitem{bao2009high}
T.~Bao, J.-D. Xu, and H.-S. Zhang, ``High-rate complex orthogonal space-time
  block codes for multiple transmit antennas,'' in \emph{Computer Science and
  Engineering, 2009. WCSE'09. Second International Workshop on}, vol.~1.\hskip
  1em plus 0.5em minus 0.4em\relax IEEE, 2009, pp. 138--141.

\bibitem{nora}
{3GPP RAN1 Meeting\#83, TR 36.859}, ``{Study on Downlink Multiuser
  Superposition Transmission (MUST) for LTE},'' Rel. 13, Nov 2015.

\end{thebibliography}

\end{document}